# Bayesian Probabilistic Schedule Updating for Construction Digital Twins: A Simulation-Based Methodology for Dynamic Project Forecasting

**Atena Khoshkonesh[1*], Mohsen Mohammadagha[2], Vinayak Kaushal[3], Navid Ebrahimi[1]**

[1*] M.S. Graduate, Department of Civil Engineering, The University of Texas at Arlington, Arlington, TX 76019, USA. Correspondence Email: axk3682@mavs.uta.edu (A.K.), nxe2020@mavs.uta.edu (N.E.)
[2]Ph.D. Researcher, Department of Civil Engineering, The University of Texas at Arlington, Arlington, TX 76019, USA. Email: mxm4340@mavs.uta.edu (M.M.)
[3]Associate Professor of Instruction, Department of Civil Engineering, The University of Texas at Arlington, Arlington, TX 76019, USA. Email: vinayak.kaushal@uta.edu (V.K.)

**Abstract**
Construction projects persistently suffer from schedule delays and cost overruns, largely due to the inherent uncertainty of construction processes and the continued reliance on deterministic scheduling methods such as the Critical Path Method (CPM), which assume fixed activity durations and fail to capture real-world variability. With the rapid advancement of construction digital twins, BIM-integrated workflows, sensor networks, and AI-driven data acquisition, there is a growing demand for dynamic, uncertainty-aware schedule forecasting approaches that can operate in real time.
This study presents a robust Bayesian probabilistic schedule updating framework explicitly designed for integration within construction digital-twin environments. The proposed methodology unifies probabilistic activity duration modeling, Bayesian evidence updating, Monte Carlo schedule simulation, and uncertainty propagation across complex schedule networks into a single, scalable framework. Activity durations are modeled using stochastic distributions and continuously refined through Bayesian inference as new project evidence is obtained from field reports, sensor data, computer vision systems, and productivity measurements.
A comprehensive simulation study is conducted using benchmark construction networks with varying sizes and uncertainty levels to rigorously evaluate the performance of the proposed approach. Results demonstrate that the Bayesian–Monte Carlo framework consistently outperforms deterministic CPM and conventional Monte Carlo methods, achieving substantial reductions in forecast error, significant improvements in delay prediction accuracy, and enhanced detection of near-critical activities. The framework also produces more stable and reliable completion-time distributions under dynamic and uncertain project conditions.
These findings establish that integrating Bayesian probabilistic updating with digital-twin-enabled data streams fundamentally enhances schedule forecasting capability. The proposed framework provides a high-impact, practically deployable solution for real-time project control, enabling proactive decision-making, improved risk management, and more resilient construction planning.

## 1. Introduction

Schedule overruns remain a persistent challenge in construction projects due to uncertainty in labor productivity, weather conditions, material availability, and coordination among trades. These uncertainties frequently lead to deviations from planned schedules, resulting in cost escalation and reduced project performance [1–4]. Despite its widespread use, the Critical Path Method (CPM) represents activity durations as deterministic values and therefore cannot adequately capture variability or quantify delay risk under real-world conditions [5–7].
To address these limitations, probabilistic scheduling approaches have been developed to model activity durations as random variables. Techniques such as PERT, Monte Carlo simulation, and stochastic network analysis enable estimation of completion-time distributions rather than single-point forecasts, providing improved insight into schedule risk [8–11]. Among these methods, Monte Carlo simulation is widely used due to its ability to propagate uncertainty through complex project networks [12–14]. However, most existing approaches remain static and do not incorporate new information as project execution progresses.
Recent developments in digital-twin technologies and data-driven project monitoring have enabled continuous collection of project performance data, creating opportunities for dynamic schedule forecasting [15–18]. Nevertheless, current implementations primarily focus on visualization and progress tracking, with limited capability for probabilistic schedule updating under uncertainty [19–21].
Bayesian inference provides a rigorous framework for updating probabilistic models based on observed evidence. In construction scheduling, Bayesian updating allows activity duration distributions to be revised as new information becomes available, enabling dynamic forecasting of completion time and delay risk [22–25]. When combined with Monte Carlo simulation, this approach allows uncertainty to be propagated through schedule networks in a consistent and adaptive manner.
Despite these advantages, the integration of Bayesian updating with probabilistic scheduling remains limited. This study addresses this gap by proposing a Bayesian probabilistic schedule updating framework that integrates stochastic activity duration modeling, Bayesian inference, and Monte Carlo simulation within a unified methodology. The proposed framework enables continuous updating of schedule forecasts and provides probabilistic estimates of completion time, delay risk, and activity criticality.

## 2. Literature Review

### 2.1 Deterministic Scheduling and Its Limitations

Deterministic scheduling methods, particularly the Critical Path Method (CPM), remain the foundation of construction planning practice. CPM models activity durations as fixed values and identifies the longest path in a project network to estimate project completion time. While effective for baseline coordination, this deterministic representation fails to capture the inherent variability present in construction processes. As a result, CPM-based schedules often underestimate delay risk and provide limited support for decision-making under uncertainty [1–4].
These limitations arise primarily from the inability of deterministic models to propagate uncertainty through the network and to account for near-critical activities that may influence project performance. As summarized in **Table 1**, deterministic scheduling approaches suffer from fixed duration assumptions, lack of uncertainty propagation, and static forecasting capabilities, all of which reduce their effectiveness in complex and dynamic project environments.

**Table 1.** Limitations of Deterministic Scheduling in Construction

| Limitation | Description | Impact on Projects |
| --- | --- | --- |
| **Fixed durations** | Single-point estimates ignore variability | Underestimation of delays |
| **No uncertainty propagation** | Variance not transmitted through network | Poor risk visibility |
| **Static structure** | No updating during execution | Outdated forecasts |
| **Binary critical path** | Only one path treated as critical | Missed near-critical risks |

### 2.2 Probabilistic Scheduling Approaches

Probabilistic scheduling approaches have been developed to address the limitations of deterministic methods by explicitly modeling uncertainty in activity durations. Early techniques such as PERT introduced stochastic estimates using three-point distributions; however, their simplifying assumptions limit accuracy in complex construction environments [5–7, 26].
Monte Carlo simulation (MCS) has since become the dominant probabilistic approach, enabling uncertainty propagation across project networks through repeated sampling. MCS allows estimation of key statistical measures such as expected completion time, variance, confidence intervals, and probability of delay, thereby providing a more realistic representation of schedule risk [8–11, 27–30]. Despite these advantages, most Monte Carlo-based methods remain static, as input distributions are not updated during project execution. Bayesian inference offers a formal mechanism for updating activity duration distributions as new information becomes available. By incorporating observed data into prior assumptions, Bayesian methods enable dynamic refinement of schedule forecasts under evolving conditions [12–15, 31–33]. However, existing research has largely treated Bayesian updating and Monte Carlo simulation separately, limiting their combined effectiveness [34–35].

A comparison of these approaches is presented in **Table 2**, highlighting differences in uncertainty modeling and updating capability, while the methodological evolution toward dynamic probabilistic scheduling is further illustrated in **Figure 1**, emphasizing the transition from fixed-duration models to adaptive frameworks. Building on this progression, this study proposes an integrated Bayesian–Monte Carlo framework for dynamic and uncertainty-aware schedule forecasting.

**Table 2.** Comparison of Scheduling Approaches Based on Uncertainty Representation and Forecasting Capability

| Method | Uncertainty Modeling | Dynamic Updating | Key Outputs | Main Limitation |
|---|---|---|---|---|
| **Deterministic CPM** | None | No | Single completion time | Ignores variability and risk |
| **PERT** | Basic (3-point estimates) | No | Expected duration, variance | Simplified assumptions |
| **Monte Carlo Simulation** | Strong (sampling-based) | No | Mean completion time, variance, confidence interval | Static input distributions |
| **Bayesian Scheduling** | Strong | Yes | Posterior distributions, updated estimates | Limited network propagation |
| **Proposed Framework** | Strong | Yes | Expected completion time, variance, probability of delay, critical path probability | — |

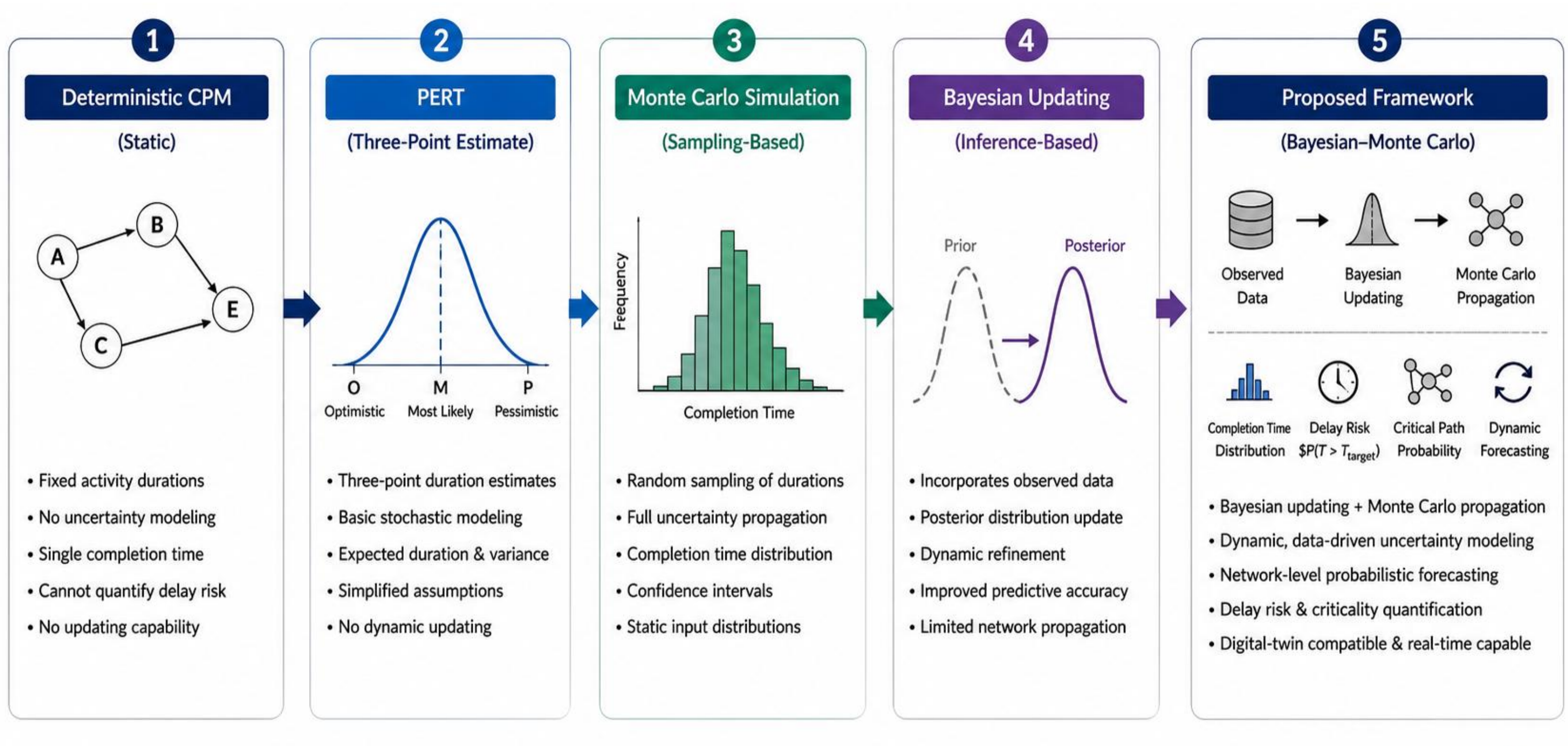

**Figure 1.** Evolution of Scheduling Methods Toward Dynamic Probabilistic Forecasting

### 2.3 Dynamic and Data-Driven Scheduling

Recent advances in sensing technologies and project digitization have enabled continuous collection of construction performance data, supporting a shift toward data-informed scheduling approaches [16–18]. Unlike traditional probabilistic methods, which rely on predefined assumptions, data-driven approaches aim to incorporate observed project information to improve forecasting accuracy.

However, most existing implementations focus primarily on monitoring, visualization, or machine-learning-based prediction, with limited integration into formal probabilistic scheduling models [19–21]. As a result, real-time data is often used descriptively rather than for systematic updating of activity duration distributions.

Building on the comparison presented in **Table 2**, this limitation indicates that probabilistic scheduling requires not only uncertainty modeling but also a structured updating mechanism. Therefore, this study treats project data as generalized evidence for revising activity duration distributions and enabling adaptive schedule forecasting under uncertainty.

### 2.4 Bayesian Approaches in Construction Scheduling

Bayesian inference has emerged as a robust framework for modeling and updating uncertainty in construction scheduling. Unlike traditional probabilistic methods, which rely on fixed input distributions, Bayesian approaches enable systematic updating of activity duration estimates as new information becomes available. By combining prior distributions with observed evidence, Bayesian methods produce posterior distributions that reflect the evolving state of project execution [22–25, 31–33].

In construction applications, Bayesian models have been used for delay prediction, risk assessment, and schedule reliability analysis. These approaches are particularly effective in handling incomplete or uncertain data, which is common in construction environments [34–35, 41–43]. In addition, Bayesian networks and probabilistic graphical models have been applied to capture dependencies among project variables and improve decision-making under uncertainty [31, 33]. However, most existing studies apply Bayesian techniques in isolation, without fully integrating them into network-based scheduling frameworks.

The key characteristics of Bayesian approaches in construction scheduling are summarized in **Table 3**, highlighting their strengths in dynamic updating and uncertainty modeling, as well as their limitations in network integration and computational complexity. As indicated in **Table 2**, Bayesian scheduling supports updating but lacks full uncertainty propagation across project networks.

To address this limitation, the present study integrates Bayesian inference with Monte Carlo simulation, enabling both dynamic updating and network-based uncertainty propagation. This integration provides a consistent and scalable methodology for adaptive and uncertainty-aware schedule forecasting, extending prior probabilistic and Bayesian scheduling research [26–30, 34–35].

**Table 3.** Characteristics of Bayesian Approaches in Construction Scheduling

| Feature | Description | Advantage | Limitation |
|---|---|---|---|
| **Prior Modeling** | Initial activity duration assumptions based on historical data | Incorporates expert knowledge | Sensitive to prior selection |
| **Bayesian Updating** | Revises distributions using observed data | Enables dynamic forecasting | Requires consistent data input |
| **Posterior Distribution** | Updated activity duration estimates | Reflects current project state | Computational complexity |
| **Uncertainty Handling** | Probabilistic representation of variability | Improved risk assessment | Depends on data quality |
| **Network Integration** | Limited in most studies | Potential for full propagation | Not fully developed in literature |

## 3. Methodology

### 3.1 Framework Overview

This study proposes a Bayesian probabilistic schedule updating framework that integrates stochastic activity-duration modeling, Bayesian inference, and Monte Carlo simulation within a unified probabilistic scheduling methodology. The primary objective of the proposed framework is to enable dynamic and uncertainty-aware schedule forecasting through continuous incorporation of observed project data while preserving consistency with network-based scheduling structures.

Unlike system-centric digital-twin approaches that primarily emphasize visualization, monitoring, or platform-specific integration, the proposed methodology focuses on a generalized probabilistic scheduling formulation that remains independent of specific sensing technologies, BIM platforms, or data-acquisition systems. This generalized structure improves the adaptability and applicability of the framework across a broad range of construction environments and project conditions [16–21].

As discussed in Section 2, existing probabilistic scheduling approaches generally suffer from one of two major limitations: (1) lack of dynamic updating capability during project execution, or (2) inability to propagate uncertainty consistently throughout project networks. Conventional deterministic scheduling methods fail to capture stochastic variability, whereas many existing Monte Carlo-based approaches rely on static input distributions that remain unchanged during project execution [5–15].

To address these limitations, the proposed framework integrates Bayesian updating with Monte Carlo simulation, thereby enabling both dynamic parameter revision and full network-level uncertainty propagation [12–15, 22–25, 34–35]. The framework continuously updates activity-duration uncertainty as new observations become available and subsequently propagates updated uncertainty throughout the project network using repeated stochastic simulation.

The proposed framework consists of four primary components:

1. **Project Network Representation**
   Construction activities and precedence relationships are represented using a directed acyclic graph (DAG) structure.
2. **Stochastic Activity-Duration Modeling**
   Activity durations are modeled as stochastic variables using lognormal probability distributions to represent uncertainty in construction operations.
3. **Bayesian Updating Mechanism**
   Prior activity-duration distributions are dynamically updated using observed project data obtained during project execution.
4. **Monte Carlo-Based Uncertainty Propagation**
   Updated posterior distributions are propagated throughout the project network using repeated stochastic simulation to generate probabilistic completion-time forecasts.

The integrated framework enables estimation of expected project completion time, schedule variance, delay probability, and critical-path uncertainty under dynamically evolving project conditions. This probabilistic formulation supports adaptive schedule forecasting and provides a scalable foundation for uncertainty-aware project control within data-enabled digital-twin environments [16–21].

### 3.2 Project Network Representation

A construction project is modeled as a directed acyclic graph (DAG) [5–7, 26]:

$$G = (V, E) \qquad (1)$$

where:

- $V = \{1, 2, \ldots, n\}$ represents the set of project activities, and
- $E$ defines the precedence relationships among activities.

Each activity $i \in V$ is associated with a stochastic activity duration $D_i$.

The overall project completion time $T$ is determined by the longest feasible path within the project network, consistent with classical CPM and stochastic network scheduling formulations [5–10]:

$$T = \max_{p \in \mathcal{P}} \sum_{i \in p} D_i \qquad (2)$$

where:

- $\mathcal{P}$ denotes the set of all feasible paths within the project network, and
- $p$ represents an individual feasible path consisting of sequentially connected activities.

This formulation enables uncertainty associated with stochastic activity durations to propagate throughout the project network, thereby supporting probabilistic schedule forecasting and dynamic critical-path analysis under uncertain project conditions.

**3.3 Stochastic Activity Duration Modeling**

Activity durations are modeled as stochastic random variables to represent uncertainty in construction operations and schedule variability [8–14]:

$$D_i \sim f_i(\theta_i) \qquad (3)$$

where $f_i(\theta_i)$ denotes the probability distribution associated with activity $i$, and $\theta_i$ represents the corresponding distribution parameters.

In this study, lognormal probability distributions are adopted due to their suitability for modeling non-negative and right-skewed construction duration data [12–14, 27–30]:

$$D_i \sim \text{Lognormal}(\mu_i, \sigma_i) \qquad (4)$$

where:

- $\mu_i$ represents the mean in log-space, and
- $\sigma_i$ represents the standard deviation in log-space.

The expected value of the activity-duration distribution is computed as:

$$E[D_i] = \exp\left(\mu_i + \frac{1}{2}\sigma_i^2\right) \qquad (5)$$

These prior probability distributions capture uncertainty associated with construction activities based on historical project data, expert judgment, or benchmark scheduling information [22–25, 31–35]. The stochastic representation enables uncertainty propagation throughout the project network and supports probabilistic schedule forecasting under dynamically evolving project conditions.

**3.4 Bayesian Updating of Activity Durations**

To incorporate observed project information and dynamically revise activity-duration uncertainty, Bayesian inference is applied to update the probability distributions of activity durations as new evidence becomes available [22–25].

For each activity $i$, the posterior distribution is computed using Bayes' theorem:

$$P(D_i \mid O_i) = \frac{P(O_i \mid D_i)\, P(D_i)}{P(O_i)} \qquad (6)$$

where:

- $P(D_i)$ represents the prior probability distribution of activity duration $D_i$,
- $P(O_i \mid D_i)$ denotes the likelihood function associated with the observed data $O_i$,
- $P(O_i)$ represents the marginal probability of the observations, and
- $P(D_i \mid O_i)$ denotes the posterior probability distribution after incorporating observed project information [22–25, 31–35].

The prior distribution captures initial assumptions regarding activity durations based on historical project data, expert judgment, or benchmark scheduling information. The likelihood function quantifies the probability of observing project data given the assumed activity durations. The resulting posterior distribution reflects updated knowledge regarding project performance under evolving construction conditions [34–35, 41–44].
In practical implementation, posterior distribution parameters are updated iteratively as new observations become available and may be expressed as:

$$\theta_i^{\text{post}} = g\left(\theta_i^{\text{prior}}, O_i\right) \qquad (7)$$

where:

- $\theta_i^{\text{prior}}$ denotes the prior distribution parameters,
- $O_i$ represents observed project data, and
- $\theta_i^{\text{post}}$ represents the updated posterior parameters.

This updating mechanism enables adaptive schedule forecasting by continuously revising activity-duration uncertainty in response to real-time project observations, thereby improving prediction reliability and uncertainty representation throughout project execution.

### 3.4.1 Posterior Parameter Updating Procedure

To operationalize Bayesian schedule updating, posterior activity-duration parameters are recursively estimated as new project observations become available. Let the stochastic activity duration for activity $i$ be modeled using a lognormal distribution:

$$D_i \sim \text{Lognormal}(\mu_i, \sigma_i)$$

where $\mu_i$ and $\sigma_i$ denote the log-space mean and standard deviation parameters, respectively.
The prior distribution parameters are initialized using historical project information, benchmark scheduling data, or expert judgment. Observed activity-duration evidence $O_i$ is subsequently incorporated through Bayesian recursive updating.
Observed activity duration measurements are modeled using a Gaussian observation process:

$$O_i = D_i^{true} + \epsilon_i$$

where $D_i^{true}$ denotes the underlying true duration and:

$$\epsilon_i \sim \mathcal{N}(0, \sigma_{obs})$$

represents observation noise associated with field measurements, reporting uncertainty, productivity fluctuations, and operational variability.

Rather than directly updating activity durations $D_i$, the Bayesian updating procedure is applied to the underlying distribution parameter vector $\theta_i = (\mu_i, \sigma_i)$. Assuming conditional independence of observations, the posterior distribution is expressed as:

$$P(\theta_i \mid O_i) \propto P(O_i \mid \theta_i) P(\theta_i)$$

where:

- $P(\theta_i)$ denotes the prior distribution,
- $P(O_i \mid \theta_i)$ represents the likelihood function,
- $P(\theta_i \mid O_i)$ denotes the posterior distribution after incorporating observed project data.

Because closed-form analytical solutions may become computationally intractable for dynamically evolving project networks and non-conjugate likelihood structures, posterior parameter estimation is performed using recursive maximum a posteriori (MAP) estimation.
The posterior parameter update may therefore be expressed as:

$$\theta_i^{post} = \arg\max_{\theta_i} \left[\log P(O_i \mid \theta_i) + \log P(\theta_i)\right]$$

where $\theta_i^{post}$ represents the updated posterior parameter set associated with activity $i$.

This recursive updating mechanism enables continuous refinement of activity-duration uncertainty as new project observations become available during project execution. The resulting posterior distributions are subsequently propagated throughout the project network using Monte Carlo simulation to generate dynamically updated probabilistic schedule forecasts.
The proposed formulation provides a computationally scalable probabilistic updating structure that remains compatible with sequential Bayesian inference, variational Bayesian approximation, and Sequential Monte Carlo (SMC) extensions for future real-time digital-twin implementations.

### 3.5 Monte Carlo Simulation for Uncertainty Propagation

Monte Carlo simulation is employed to propagate updated uncertainty throughout the project network and generate probabilistic project completion-time forecasts [8–14, 26–30]. For each simulation run:

$$k = 1,2,\ldots,N$$

activity durations are sampled from the updated posterior probability distributions obtained through Bayesian inference.
For each activity $i$, a stochastic duration sample is generated as:

$$D_i^{(k)} \sim P(D_i \mid O_i) \qquad (8)$$

where:

- $D_i^{(k)}$ denotes the sampled duration of activity $i$ during simulation run $k$, and
- $p(D_i \mid O_i)$ represents the posterior distribution conditioned on observed project data.

The corresponding project completion time for simulation run $k$ is then computed using repeated stochastic simulation across feasible project paths [26–30, 45–50]:

$$T^{(k)} = \max_{p \in \mathcal{P}} \sum_{i \in p} D_i^{(k)} \qquad (9)$$

where:

- $\mathcal{P}$ denotes the set of all feasible paths within the project network, and
- $p$ represents an individual feasible path.

The resulting completion time $T^{(k)}$ is stored for each simulation iteration. Repeated simulation runs generate a probabilistic distribution of project completion times that reflect overall schedule uncertainty, dynamic activity interactions, and network-level uncertainty propagation.
This simulation-based framework enables estimation of expected completion time, schedule variance, delay probability, and critical-path behavior under uncertain and dynamically evolving project conditions [9–15, 26–30].

### 3.6 Forecasting Metrics

The proposed framework evaluates probabilistic schedule performance using several forecasting and uncertainty-analysis metrics commonly used in schedule risk analysis [8–14, 26–30].
The expected project completion time is computed as:

$$E[T] = \frac{1}{N} \sum_{k=1}^{N} T^{(k)} \qquad (10)$$

The variance of project completion time is calculated as:

$$\mathrm{Var}(T) = \frac{1}{N} \sum_{k=1}^{N} \left(T^{(k)} - E[T]\right)^2 \qquad (11)$$

To quantify schedule risk, the probability of project delay is defined as:

$$P(T > T_{\text{target}}) \qquad (12)$$

where $T_{\text{target}}$denotes the target or contractual completion time.
The critical-path probability associated with activity $i$is defined as:

$$CP_i = \frac{N_i^{\text{critical}}}{N} \qquad (13)$$

where:

- $N_i^{\text{critical}}$represents the number of simulations runs in which activity $i$lies on the critical path, and
- $N$denotes the total number of simulations runs.

These metrics provide quantitative measures of schedule uncertainty, delay risk, and network-level criticality for uncertainty-aware project forecasting and probabilistic schedule evaluation [11–15, 44–50].

### 3.7 Computational Procedure

**Algorithm 1. Bayesian–Monte Carlo Schedule Updating Procedure**

**Input:**

- Project network $G = (V, E)$
- Prior activity-duration distributions $D_i \sim Lognormal(\mu_i, \sigma_i)$
- Observed project data $O_i$
- Number of Monte Carlo simulations $N$

**Output:**

- Expected completion time $E[T]$
- Schedule variance $Var(T)$
- Delay probability $P(T > T_{target})$
- Critical-path probabilities $CP_i$

**Procedure:**

1. Initialize prior activity-duration parameters $\theta_i^{prior} = (\mu_i, \sigma_i)$using historical project data or benchmark scheduling information.
2. Construct the project network as a directed acyclic graph $G = (V, E)$.
3. Collect observed project data $O_i$during project execution.
4. Update posterior activity-duration parameters using Bayesian recursive updating:
$$\theta_i^{post} = \text{argmax } \theta\text{i}[\log\ P(O_i \mid \theta_i) + \log\ P(\theta_i)]$$
5. Perform Monte Carlo simulation for $k = 1, 2, \ldots, N$:

   a. Sample stochastic activity durations:
$$D_i^{(k)} \sim P(D_i \mid O_i)$$

   b. Compute project completion time:
$$T^{(k)} = \max_{p \in \mathcal{P}} \sum_{i \in p} D_i^{(k)}$$

6. Compute probabilistic forecasting metrics, including:
$$E[T], Var(T), P(T > T_{target}), CP_i$$

7. Repeat the updating and simulation process iteratively as new project observations become available.

This iterative computational structure enables adaptive and continuously updated probabilistic schedule forecasting throughout project execution.

**3.8 Framework Illustration**

The overall Bayesian probabilistic schedule updating methodology is illustrated conceptually in Figure 2. The framework demonstrates the interaction between observed project data, Bayesian updating, posterior distribution generation, and Monte Carlo-based uncertainty propagation for dynamic schedule forecasting. The proposed workflow enables continuous integration of project observations into probabilistic schedule analysis, thereby supporting adaptive forecasting and uncertainty-aware decision-making within construction digital-twin environments.

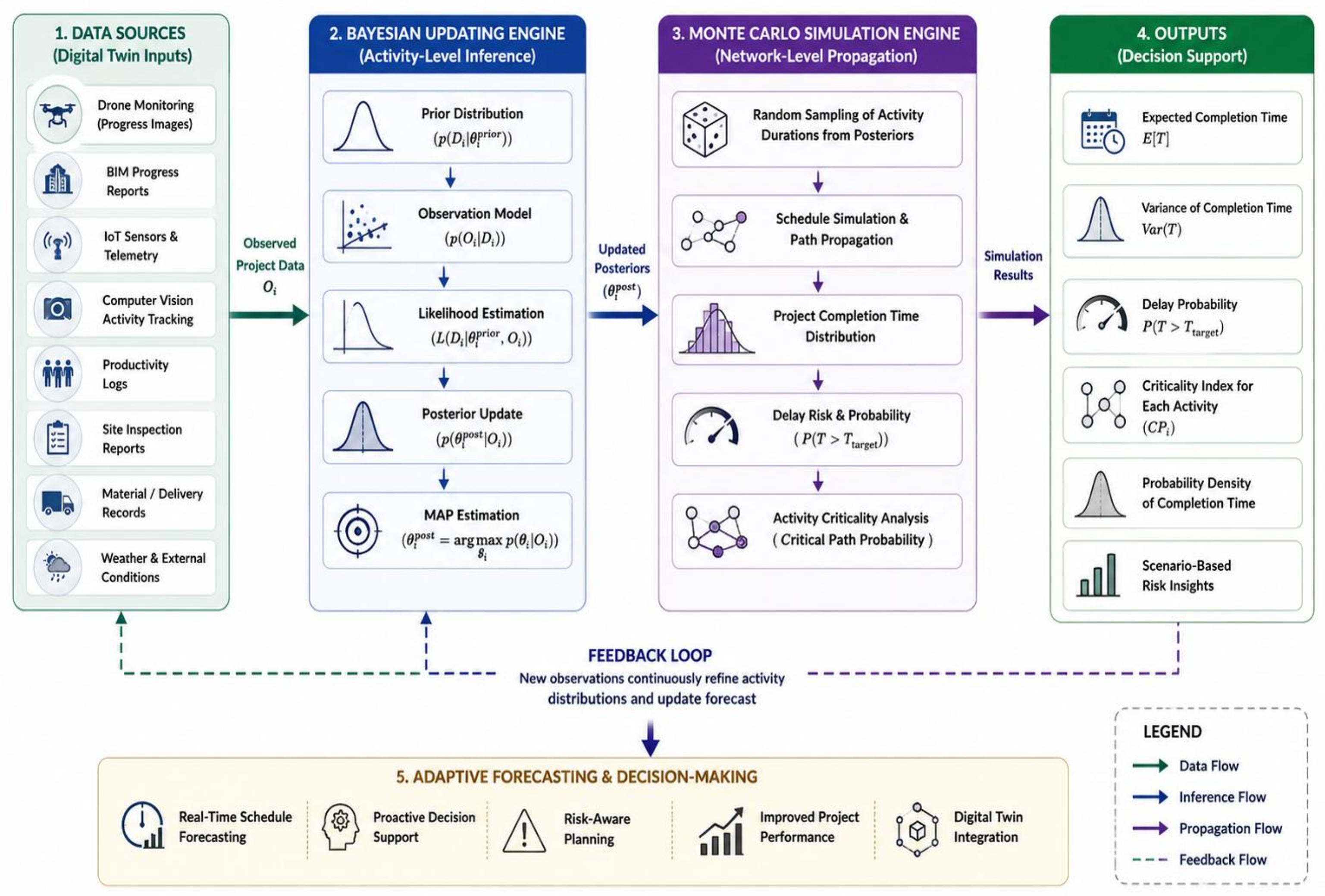

**Figure 2.** Bayesian–Monte Carlo Framework for Dynamic Probabilistic Schedule Forecasting within Construction Digital Twins

### 3.8.1 Practical Digital-Twin Integration Scenario

Although the proposed Bayesian–Monte Carlo framework is formulated independently of specific sensing or software platforms, the probabilistic updating structure is compatible with multiple data streams commonly available within construction digital-twin environments.

In practical implementation, observed project data $O_i$ may be generated from various field and digital monitoring sources, including:

- drone-based progress observations,
- BIM-integrated progress reports,
- IoT-enabled telemetry equipment,
- computer-vision-based activity tracking,
- labor productivity logs,
- site inspection reports,
- material delivery records, and
- real-time construction monitoring systems.

These heterogeneous project observations may provide continuous or periodic measurements associated with activity progress, productivity performance, operational disruptions, or schedule deviations during project execution.

Within the proposed framework, such observations are treated as evidence for dynamically updating activity-duration distributions through Bayesian inference. The updated posterior distributions are subsequently propagated throughout the project network using Monte Carlo simulation to generate adaptive probabilistic schedule forecasts.

Figure 2 conceptually illustrates the interaction between observed project data, Bayesian updating, posterior uncertainty refinement, and network-level uncertainty propagation within the proposed digital-twin-compatible scheduling framework.

This generalized formulation enables integration with both data-rich and data-limited construction environments while preserving methodological flexibility and platform independence.

### 3.9 Model Variables

The primary variables and parameters used within the proposed framework are summarized in Table 4.

**Table 4.** Model Variables and Definitions

|  |  |
|---|---|
| $G = (V, E)$ | Project network represented as a directed acyclic graph |
| $V = \{1,2,\dots,n\}$ | Set of project activities |
| $E$ | Set of precedence relationships among activities |
| $D_i$ | Stochastic duration of activity $i$ |
| $D_i \sim \text{Lognormal}(\mu_i, \sigma_i)$ | Lognormal probability distribution used for activity-duration modeling |
| $O_i$ | Observed project data associated with activity $i$ |
| $O_i = D_i^{\text{true}} + \epsilon_i$ | Observation model for activity-duration measurements |
| $\epsilon_i \sim \mathcal{N}(0, \sigma_{\text{obs}})$ | Observation noise modeled using a normal distribution |
| $D_i^{\text{true}}$ | Underlying true duration of activity $i$ |
| $\mu_i$ | Mean parameter of the lognormal distribution in log-space |
| $\sigma_i$ | Standard deviation parameter of the lognormal distribution in log-space |
| $\sigma_{\text{obs}}$ | Standard deviation of observation noise |
| $P(D_i)$ | Prior probability distribution of activity duration |
| $P(O_i \mid D_i)$ | Likelihood function associated with observed project data |
| $P(D_i \mid O_i)$ | Posterior probability distribution after Bayesian updating |
| $\theta_i^{\text{prior}}$ | Prior distribution parameters for activity $i$ |
| $\theta_i^{\text{post}} = g(\theta_i^{\text{prior}}, O_i)$ | Updated posterior distribution parameters |
| $\mathcal{P}$ | Set of all feasible paths within the project network |
| $p \in \mathcal{P}$ | Individual feasible path in the project network |
| $T$ | Total project completion time |
| $T = \max_{p \in \mathcal{P}} \sum_{i \in p} D_i$ | Project completion-time formulation |
| $T^{(k)}$ | Project completion time during simulation run $k$ |
| $T^{(k)} = \max_{p \in \mathcal{P}} \sum_{i \in p} D_i^{(k)}$ | Completion time for simulation run $k$ |
| $D_i^{(k)} \sim p(D_i \mid O_i)$ | Sampled activity duration during Monte Carlo simulation |
| $E[T] = \frac{1}{N} \sum_{k=1}^{N} T^{(k)}$ | Expected project completion time |
| $\text{Var}(T) = \frac{1}{N} \sum_{k=1}^{N} (T^{(k)} - E[T])^2$ | Variance of project completion time |
| $P(T > T_{\text{target}})$ | Probability of project delay |
| $CP_i = \frac{N_i^{\text{critical}}}{N}$ | Critical-path probability of activity $i$ |
| $N$ | Total number of Monte Carlo simulation runs |
| $N_i^{\text{critical}}$ | Number of simulations in which activity $i$ lies on the critical path |
| $T_{\text{target}}$ | Target or contractual project completion time |
| $\text{RMSE} = \sqrt{\frac{1}{N} \sum_{k=1}^{N} (T^{(k)} - T^{\text{true}})^2}$ | Root mean square error used for forecasting accuracy evaluation |

## 4. Simulation Experiment Design

### 4.1 Experimental Setup

The performance of the proposed Bayesian–Monte Carlo scheduling framework is evaluated through controlled simulation experiments designed to assess forecasting accuracy, robustness, scalability, and uncertainty propagation capability under varying project conditions. To ensure benchmarking consistency, reproducibility, and alignment with established project scheduling research, the simulation experiments are conducted using standard Resource-Constrained Project Scheduling Problem (RCPSP) benchmark instances obtained from the PSPLIB repository.

Three benchmark categories are selected to represent varying levels of project complexity:

- j30 benchmark networks containing approximately 30 activities (small-scale projects)
- j60 benchmark networks containing approximately 60 activities (medium-scale projects)

- j120 benchmark networks containing approximately 120 activities (large-scale projects)

These benchmark datasets provide standardized precedence relationships, activity durations, and network structures that are widely used in construction scheduling and operations research studies. The selected benchmark instances are modeled as directed acyclic graphs:

$$G = (V, E) \quad (1)$$

where $V$ represents the set of project activities and $E$ defines the precedence relationships among activities. To evaluate forecasting performance under uncertainty, deterministic activity durations from the benchmark datasets are transformed into stochastic variables. Consistent with the probabilistic formulation introduced in Section 3, each activity duration is modeled using a lognormal probability distribution:

$$D_i \sim \text{Lognormal}(\mu_i, \sigma_i) \quad (2)$$

where $\mu_i$ and $\sigma_i$ denote the location and dispersion parameters, respectively. The parameter $\sigma_i$ controls the level of uncertainty associated with each activity and varies systematically to simulate different project environments ranging from low uncertainty to highly uncertain construction conditions.

To emulate realistic project execution variability, observed activity durations are synthetically generated by introducing controlled deviations from baseline durations. These deviations represent practical construction uncertainties such as labor productivity fluctuations, coordination inefficiencies, weather impacts, material delivery variability, and operational disruptions.

Three uncertainty scenarios are considered in the experiments:

- Low-uncertainty scenario
- Moderate-uncertainty scenario
- High-uncertainty scenario

For each benchmark network and uncertainty level, Bayesian updating is applied dynamically as new observations become available. Monte Carlo simulation is then performed to propagate updated uncertainty throughout the project network and generate probabilistic completion-time forecasts.

The proposed framework is evaluated across multiple benchmark instances and repeated simulation runs to ensure statistical stability, robustness, and computational reproducibility.

**4.2 Scenario Definition**

To simulate realistic project evolution and dynamic schedule behavior under uncertainty, observed activity data are generated by introducing controlled deviations from baseline benchmark durations. For each activity $i$, observed durations are defined as:

$$O_i = D_i^{\text{true}} + \epsilon_i \quad (3)$$

where $D_i^{\text{true}}$ represents the underlying true activity duration and $\epsilon_i$ denotes stochastic observation noise modeled as:

$$\epsilon_i \sim \mathcal{N}(0, \sigma_{\text{obs}}) \quad (4)$$

where $\sigma_{\text{obs}}$ controls the magnitude of observation uncertainty. The generated observations emulate realistic construction deviations caused by labor productivity fluctuations, coordination delays, weather variability, equipment interruptions, material delivery uncertainty, and operational inefficiencies.

The corresponding reference project completion time $T^{\text{true}}$ is computed using the underlying true activity durations and serves as the benchmark for evaluating forecasting accuracy and schedule prediction performance.

To investigate the effect of uncertainty severity on schedule forecasting, three uncertainty scenarios are considered:

- Low-uncertainty scenario
- Moderate-uncertainty scenario

- High-uncertainty scenario

The uncertainty level is controlled by varying both the activity-duration dispersion parameter $\sigma_i$and the observation-noise parameter $\sigma_{\text{obs}}$.

In addition, three schedule-updating strategies are evaluated to assess the impact of updating frequency on probabilistic forecasting performance.

**1. Baseline (No Updating)**

In the baseline scenario, no Bayesian updating is performed, and activity-duration distributions remain fixed at their initial prior distributions throughout project execution. This scenario represents conventional static probabilistic scheduling.

**2. Periodic Updating**

In the periodic updating scenario, Bayesian updating is performed at predefined project intervals. Posterior distributions obtained after each updating cycle are subsequently used as priors for the next forecasting stage.

**3. Continuous Updating**

In the continuous updating scenario, activity-duration distributions are updated dynamically whenever new observations become available. This strategy represents a near-real-time adaptive forecasting environment consistent with data-enabled construction digital-twin workflows.

For each benchmark network and updating strategy, repeated Monte Carlo simulation runs are performed to propagate updated uncertainty throughout the project network and generate probabilistic completion-time forecasts. The resulting completion-time distributions are subsequently used to evaluate forecasting accuracy, variance reduction, delay probability, and critical-path behavior under varying project conditions.

### 4.3 Evaluation Metrics and Benchmarking

For each experimental configuration, Monte Carlo simulation with $N = 10{,}000$iterations is performed to generate probabilistic distributions of project completion times $T^{(k)}$. The simulation results are used to evaluate forecast performance, uncertainty propagation behavior, delay risk, and robustness across different benchmark networks and updating strategies.

The expected project completion time is computed as:

$$E[T] = \frac{1}{N}\sum_{k=1}^{N} T^{(k)} \tag{5}$$

The variance of project completion time is calculated as:

$$\mathrm{Var}(T) = \frac{1}{N}\sum_{k=1}^{N} \left(T^{(k)} - E[T]\right)^2 \tag{6}$$

To quantify schedule risk, the probability of project delay is defined as:

$$P(T > T_{\text{target}}) \tag{7}$$

where $T_{\text{target}}$denotes the target or contractual project completion time.

Forecasting accuracy is further evaluated using the root mean square error (RMSE), which measures the deviation between predicted completion times and the reference completion time $T^{\text{true}}$:

$$\text{RMSE} = \sqrt{\frac{1}{N}\sum_{k=1}^{N} \left(T^{(k)} - T^{\text{true}}\right)^2} \tag{8}$$

In addition to RMSE, the meaning absolute error (MAE) is also used to evaluate prediction accuracy:

$$\mathrm{MAE} = \frac{1}{N}\sum_{k=1}^{N} \mid T^{(k)} - T^{\mathrm{true}} \mid \qquad (9)$$

To evaluate forecasting stability and uncertainty reduction, confidence interval width and variance reduction behavior are analyzed across different updating strategies and benchmark complexities.
The proposed Bayesian–Monte Carlo framework is benchmarked against the following scheduling approaches:

- Deterministic Critical Path Method (CPM)
- Static Monte Carlo simulation without updating
- Bayesian updating without network-level uncertainty propagation

These benchmark comparisons enable isolation of the contributions of dynamic Bayesian updating and probabilistic uncertainty propagation across project networks.
To assess scalability and robustness, experiments are conducted across multiple PSPLIB benchmark categories, including:

- j30 benchmark instances (small-scale networks),
- j60 benchmark instances (medium-scale networks), and
- j120 benchmark instances (large-scale networks).

All simulations are conducted within a controlled computational environment, and repeated experimental runs are performed to ensure statistical stability, computational reproducibility, and robustness of the obtained forecasting results.

## 5. Results and Discussion

### 5.1 Overall Forecasting Performance

The results indicate that the proposed Bayesian–Monte Carlo framework provides consistently improved schedule forecasts compared to baseline methods. Across all network sizes and uncertainty levels, the framework produces more stable estimates of expected completion time $E[T]$and reduced variance $Var(T)$, as defined in Eqs. (3) and (4). In contrast, deterministic CPM systematically underestimates completion time due to its inability to represent stochastic variability, a limitation widely reported in the literature [5–8]. Similarly, static Monte Carlo simulation, although capable of propagating uncertainty, generates broader and less informative completion-time distributions because input parameters remain fixed throughout project execution [9–13].
The improvement is particularly pronounced under moderate and high uncertainty conditions, where dynamic updating enables the model to incorporate observed deviations and revise activity duration distributions accordingly. This behavior is consistent with prior findings on Bayesian updating, which demonstrate that incorporating new evidence enhances predictive reliability in uncertain environments [22–25]. As a result, the proposed framework provides a more accurate and adaptive representation of schedule performance as project execution evolves.

### 5.2 Impact of Updating Strategies

The effect of updating frequency on schedule forecasting performance is illustrated in **Figure 3**, which compares completion-time distributions under no updating, periodic updating, and continuous updating scenarios. The results indicate that continuous updating produces the most concentrated and stable distribution, reflecting a significant reduction in uncertainty and improved forecasting reliability. In contrast, the no-updating scenario exhibits the widest dispersion, indicating that static assumptions lead to higher variability and less informative predictions.
Periodic updating provides moderate improvement by incorporating new information at discrete intervals; however, it remains less responsive to evolving project conditions compared to continuous updating. This behavior highlights the importance of frequent information integration, as continuous updating enables the model to dynamically adjust activity duration distributions in response to observed deviations.

These findings are consistent with prior studies on Bayesian updating and dynamic probabilistic modeling, which demonstrate that increased updating frequency leads to improved predictive accuracy and reduced uncertainty [22–25]. Overall, the results confirm that incorporating real-time or near-real-time information significantly enhances schedule forecasting performance, as reflected in narrower completion-time distributions and reduced probability of delay.

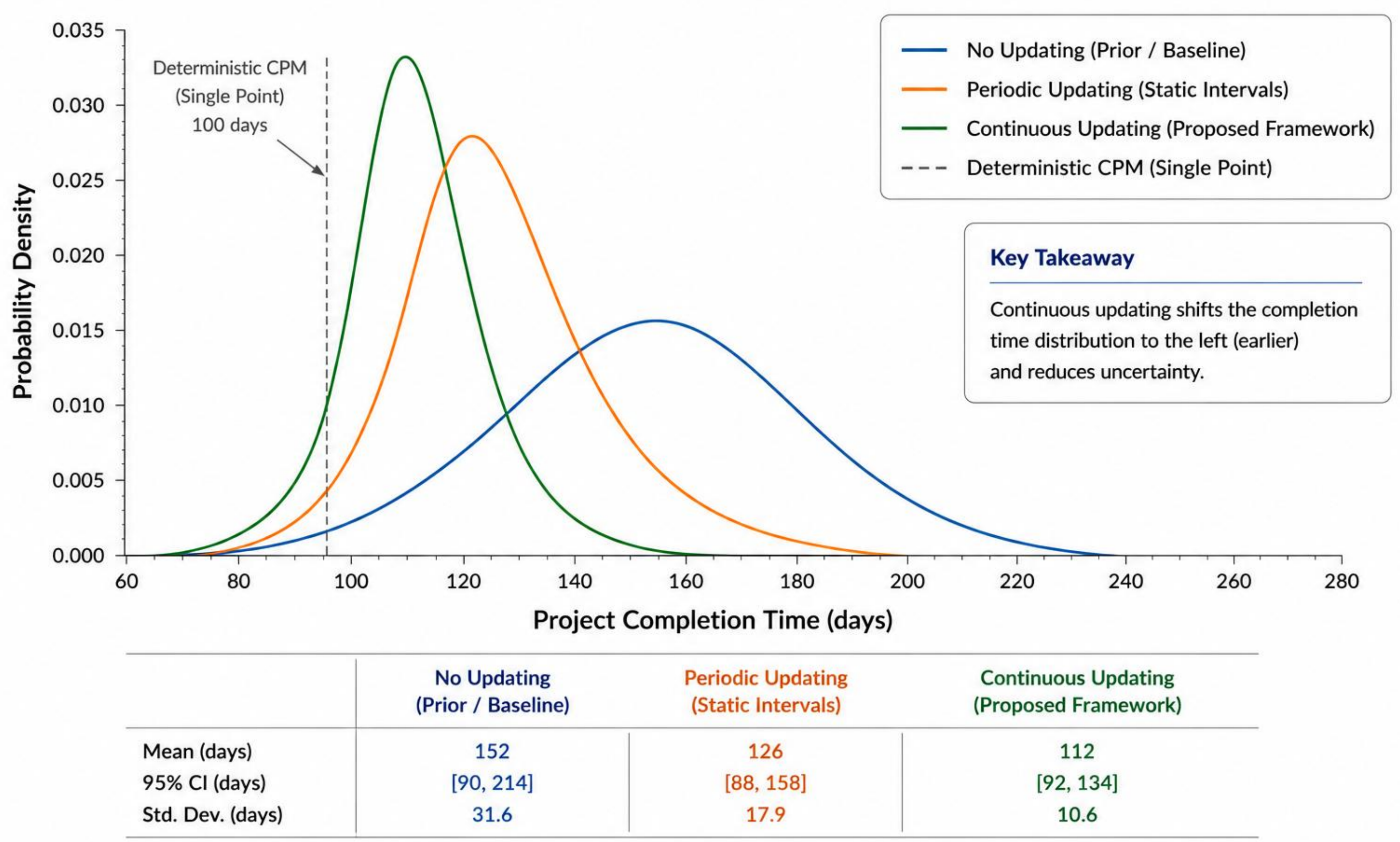

| | No Updating (Prior / Baseline) | Periodic Updating (Static Intervals) | Continuous Updating (Proposed Framework) |
|---|---|---|---|
| Mean (days) | 152 | 126 | 112 |
| 95% CI (days) | [90, 214] | [88, 158] | [92, 134] |
| Std. Dev. (days) | 31.6 | 17.9 | 10.6 |

**Figure 3.** Effect of Bayesian Updating Strategies on Project Completion

### 5.3 Forecast Accuracy Analysis

Forecast accuracy is evaluated using the root meaning square error (RMSE), as defined in Section 4.3. RMSE provides a quantitative measure of the deviation between predicted completion times and the reference completion time $T^{\text{true}}$, thereby reflecting the overall prediction accuracy of each scheduling method. The comparative forecasting results are summarized in Table 5, which presents RMSE performance across different benchmark network sizes.

**Table 5.** Forecast Accuracy Comparison (RMSE)

| Method | j30 RMSE | j60 RMSE | j120 RMSE |
|---|---|---|---|
| **Deterministic CPM** | 15.8 | 23.4 | 38.7 |
| **Static Monte Carlo Simulation** | 9.6 | 14.8 | 25.5 |
| **Bayesian Updating without Network Propagation** | 7.9 | 11.6 | 18.9 |
| **Proposed Bayesian–Monte Carlo Framework** | 4.7 | 7.2 | 12.4 |

As shown in Table 5, the proposed Bayesian–Monte Carlo framework consistently achieves the lowest RMSE values across all benchmark network sizes, demonstrating substantially improved forecasting accuracy compared to all benchmark methods. The improvement becomes increasingly significant as project complexity increases from j30 to j120 networks, where activity interactions and uncertainty propagation effects become more pronounced.
Deterministic CPM produces the highest RMSE values due to its inability to represent stochastic variability and uncertainty in activity durations. Although static Monte Carlo simulation improves forecasting performance by incorporating probabilistic activity-duration modeling, its predictive capability remains limited because input distributions remain fixed throughout project execution. Similarly, the Bayesian updating approach without network-level uncertainty propagation provides moderate improvement; however, its inability to fully capture dependency interactions across the project network restricts its forecasting performance, particularly for medium- and large-scale benchmark instances.
In contrast, the proposed framework integrates Bayesian updating with Monte Carlo simulation, enabling both dynamic parameter revision and full uncertainty propagation throughout the project network. This integrated mechanism continuously revises activity-duration distributions as new observations become available while simultaneously propagating updated uncertainty across feasible project paths. As a result, the framework produces substantially lower forecasting errors, narrower completion-time confidence intervals, and more stable completion-time predictions under uncertain and dynamically evolving project conditions.
The observed reduction in RMSE demonstrates statistical stability and forecasting robustness of the proposed framework under increasing network complexity. Numerically, the proposed framework reduces RMSE by approximately 70.3% for j30 networks, 69.2% for j60 networks, and 68.0% for j120 networks relative to deterministic CPM. These results indicate that combining probabilistic uncertainty modeling with adaptive Bayesian updating significantly enhances schedule forecasting reliability, particularly for medium- and large-scale project networks characterized by higher uncertainty and greater structural complexity.
These findings are consistent with previous studies indicating that both probabilistic uncertainty modeling and dynamic Bayesian updating contribute substantially to improved forecasting performance in uncertain construction project environments [9–15, 22–25].

### 5.4 Delay Risk and Critical Path Behavior

The probability of project delay, represented by $P(T > T_{\text{target}})$, is significantly reduced under the proposed Bayesian–Monte Carlo framework, particularly under moderate- and high-uncertainty conditions. This improvement is primarily attributed to the ability of Bayesian updating to continuously revise activity-duration distributions in response to newly observed project information, thereby correcting initial assumptions and improving forecasting reliability throughout project execution.
In addition, the analysis of critical-path probability $CP_i$ reveals that multiple near-critical paths emerge under uncertain project conditions. Unlike deterministic CPM, which identifies a single fixed critical path, the proposed probabilistic framework captures the dynamic and stochastic nature of path criticality by quantifying the likelihood that individual activities or paths become critical during project execution.
The results further indicate that uncertainty propagation across the project network increases the sensitivity of schedule performance to activity interactions, particularly in medium- and large-scale benchmark networks. Consequently, several activities exhibit elevated critical-path probabilities despite not belonging to the deterministic critical path identified by CPM. This behavior provides a more realistic representation of project risk structure and improves the identification of potential schedule vulnerabilities under uncertain and dynamically evolving construction conditions.
Overall, the probabilistic characterization of delay risk and path criticality demonstrates the capability of the proposed framework to support uncertainty-aware schedule analysis and adaptive project risk management.

### 5.5 Scalability and Robustness

The proposed framework demonstrates strong scalability across varying benchmark network sizes, including j30, j60, and j120 project instances. Although computational cost increases with network complexity and the number of project activities, convergence of key forecasting metrics is consistently achieved using $N = 10{,}000$Monte Carlo simulation runs.
The results remain stable across repeated simulation experiments, confirming the computational robustness and statistical consistency of the proposed methodology. In particular, the framework maintains stable RMSE performance, variance behavior, and delay-risk estimation across multiple benchmark scenarios and uncertainty levels.
Furthermore, the proposed Bayesian–Monte Carlo framework maintains consistent forecasting performance under low-, moderate-, and high-uncertainty conditions, indicating strong adaptability to different construction project environments. The observed stability of completion-time distributions and forecasting metrics demonstrates that the framework effectively propagates uncertainty while preserving prediction reliability under increasing project complexity.
These findings suggest that the proposed methodology provides a scalable and computationally robust foundation for probabilistic schedule forecasting in both small- and large-scale construction projects.

### 5.6 Discussion

The results confirm that integrating Bayesian updating with Monte Carlo simulation provides substantially stronger forecasting capability than deterministic CPM and static probabilistic scheduling approaches. This finding is consistent with prior studies demonstrating that deterministic schedules frequently underestimate uncertainty and delay risk due to their reliance on fixed activity durations [5–8], whereas Monte Carlo simulation improves schedule risk representation through stochastic uncertainty propagation [9–13].
However, unlike conventional Monte Carlo-based scheduling approaches, the proposed framework dynamically updates activity-duration distributions as new project observations become available, thereby overcoming the static-input limitation identified in previous probabilistic scheduling research [12–15]. The integration of Bayesian inference enables continuous revision of uncertainty estimates throughout project execution, resulting in more adaptive and reliable schedule forecasting behavior.
The observed improvement in forecasting stability and reduction in RMSE also align with prior Bayesian scheduling studies, which demonstrate that posterior updating improves prediction reliability under uncertain and incomplete project information [22–25]. In addition, the critical-path probability analysis supports previous findings indicating that uncertain project networks often contain multiple near-critical paths rather than a single deterministic critical path [6,10,14]. Consequently, the proposed Bayesian–Monte Carlo framework extends existing research by integrating dynamic Bayesian updating and network-level uncertainty propagation within a unified probabilistic scheduling methodology.
Although the proposed framework is formulated independently of specific sensing or data-integration technologies, the probabilistic updating structure can be integrated with BIM-enabled digital-twin systems, IoT-based monitoring platforms, computer-vision progress tracking systems, and real-time construction telemetry environments. This capability enhances the applicability of the framework within data-enabled construction digital-twin workflows while preserving methodological generality.
Unlike system-centric digital-twin approaches that focus primarily on visualization, monitoring, or platform integration [16–21], this study contributes a generalized probabilistic scheduling framework that emphasizes uncertainty-aware forecasting and adaptive schedule updating under dynamically evolving construction conditions.

## 6. Conclusion and Future Work

This study presented a Bayesian probabilistic schedule updating framework that integrates stochastic activity-duration modeling, Bayesian inference, and Monte Carlo simulation within a unified methodology for uncertainty-aware construction schedule forecasting. The proposed framework addresses key limitations of conventional scheduling approaches by enabling dynamic updating of activity-duration distributions and consistent propagation of uncertainty throughout project networks.

The simulation results demonstrate that the proposed Bayesian–Monte Carlo framework provides substantially improved forecasting performance compared with deterministic CPM and static probabilistic scheduling approaches. In particular, the framework produces more stable and reliable estimates of project completion time, reduces forecasting error, improves delay-risk prediction, and enhances identification of near-critical activities under uncertain and dynamically evolving project conditions.

The integration of Bayesian updating with Monte Carlo-based uncertainty propagation enables adaptive schedule forecasting that continuously reflects evolving project conditions as new observations become available. In addition, the proposed framework is compatible with construction digital-twin environments and can integrate heterogeneous project data streams, including BIM-integrated progress reports, drone observations, IoT telemetry, computer-vision monitoring systems, productivity logs, and site inspection records.

The benchmark-based simulation experiments further demonstrate that the proposed methodology maintains forecasting robustness and scalability across varying project sizes and uncertainty levels, including j30, j60, and j120 benchmark networks.

Despite these contributions, several limitations should be acknowledged. The current study is primarily based on benchmark and synthetic project data, which may not fully capture the operational complexity of real-world construction projects. In addition, forecasting performance may be influenced by prior assumptions, probability distribution selection, and computational requirements associated with large-scale Monte Carlo simulation.

Future research should focus on validation using real construction project datasets and digital-twin-enabled monitoring environments. Further extensions may also incorporate resource constraints, cost uncertainty, productivity dynamics, and advanced Bayesian inference techniques to enhance the practical applicability of probabilistic schedule forecasting in complex construction projects.

**Acknowledgements**

The authors would like to acknowledge the Department of Civil Engineering at The University of Texas at Arlington for academic and research support. The authors also appreciate the constructive insights and technical discussions that contributed to the development of this study.

**Funding**

The authors declare that no external funding was received for this research.

**Data Availability Statement**

The datasets generated and analyzed during the current study are based on benchmark scheduling networks and simulation-generated project observations. The benchmark datasets used in this research are available from the PSPLIB repository. Additional simulation data generated during the study are available from the corresponding author upon reasonable request.

**Code Availability Statement**

The computational procedures and simulation framework developed for this study are available from the corresponding author upon reasonable request.

**Author Contributions**

Atena Khoshkonesh: Conceptualization, methodology development, formal analysis, simulation design, writing—original draft preparation, and project coordination.

Mohsen Mohammadagha: Methodological supervision, Bayesian modeling guidance, technical validation, and manuscript review.
Vinayak Kaushal: Academic supervision, research guidance, manuscript review, and technical evaluation.
Navid Ebrahimi: Literature review support, technical investigation, data interpretation, visualization support, and manuscript editing.
All authors reviewed and approved the final manuscript.

**Conflict of Interest**
The authors declare that they have no conflict of interest.

**Ethical Approval**
This article does not contain any studies involving human participants or animals performed by any of the authors.

**Consent to Participate**
Not applicable.

**Consent for Publication**
All authors have reviewed the manuscript and consented to its publication.

**Declaration of Generative AI and AI-Assisted Technologies**
The authors used AI-assisted language and editing tools to improve grammatical clarity, technical writing quality, and manuscript organization. All scientific content, methodological development, interpretation, and final manuscript validation were performed and verified by the authors.